# A Calculus for Generating Ground Explanations* (Technical Report)

Mnacho Echenim, Nicolas Peltier
University of Grenoble†
LIG, Grenoble INP/CNRS

**Abstract**

We present a modification of the superposition calculus that is meant to generate explanations why a set of clauses is satisfiable. This process is related to abductive reasoning, and the explanations generated are clauses constructed over so-called abductive constants. We prove the correctness and completeness of the calculus in the presence of redundancy elimination rules, and develop a sufficient condition guaranteeing its termination; this sufficient condition is then used to prove that all possible explanations can be generated in finite time for several classes of clause sets, including many of interest to the SMT community. We propose a procedure that generates a set of explanations that should be useful to a human user and conclude by suggesting several extensions to this novel approach.

## 1 Introduction

The verification of complex systems is generally based on proving the validity, or, dually, the satisfiability of a logical formula. The standard practice consists in translating the behavior of the system to be verified into a logical formula, and proving that the negation of the formula is unsatisfiable. These formulas may be domain-specific, so that it is only necessary to test the satisfiability of the formula modulo some background theory, whence the name *Satisfiability Modulo Theories problems*, or *SMT problems*. If the formula is actually satisfiable, this means the system is not error-free, and any model can be viewed as a trace that generates an error. The models of a satisfiable formula can therefore help the designers of the system guess the origin of the errors and deduce how they can be corrected. Yet, this still requires some work. Indeed, there are generally many interpretations on different domains that satisfy the formula, and it is necessary to further analyze these models to understand where the error(s) may come from.

We present what is, to the best of our knowledge, a novel approach to this debugging problem: we argue that rather than studying one model of a formula, more valuable

*This work has been partly funded by the project ASAP of the French *Agence Nationale de la Recherche* (ANR-09-BLAN-0407-01).

†emails: Mnacho.Echenim@imag.fr, Nicolas.Peltier@imag.fr

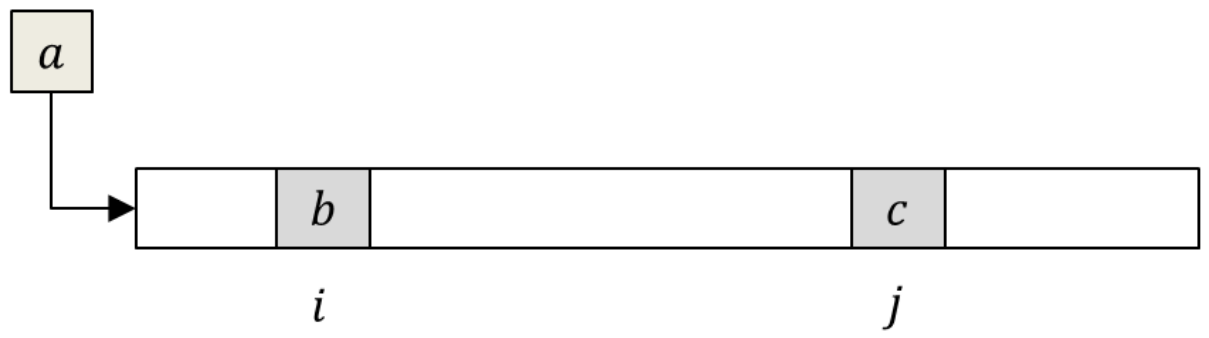

Figure 1: Insertion into array $a$ of element $b$ at position $i$ and element $c$ at position $j$.

information can be extracted from the properties that hold in *all* the models of the formula. For instance, consider the theory of arrays, which is axiomatized as follows (as introduced by [12]):

$$\forall x, z, v.\ \text{select}(\text{store}(x, z, v), z) \simeq v, \quad (1)$$

$$\forall x, z, w, v.\ z \simeq w \vee \text{select}(\text{store}(x, z, v), w) \simeq \text{select}(x, w). \quad (2)$$

These axioms state that if element $v$ is inserted into array $x$ at position $z$, then the resulting array contains $v$ at position $z$, and the same elements as in $x$ elsewhere. Assume that to verify that the order in which elements are inserted into a given array does not matter, the satisfiability of the following formula is tested (see also Figure 1):

$$\text{select}(\text{store}(\text{store}(a, i, b), j, c), k) \not\simeq \text{select}(\text{store}(\text{store}(a, j, c), i, b), k).$$

This formula asserts that there is a position $k$ that holds different values in the array obtained from $a$ by first inserting element $b$ at position $i$ and then element $c$ at position $j$, and in the array obtained from $a$ by first inserting element $c$ at position $j$ and then element $b$ at position $i$. It turns out that this formula is actually satisfiable, which in this case means that some hypotheses are missing. State of the art SMT solvers such as Yices [14] can help find out what hypotheses are missing by outputting a model of the formula. In this case, Yices outputs `(= b 1) (= c 3) (= i 2) (= k 2) (= j 2)`, and for this simple example, such a model may be sufficient to quickly understand where the error comes from. However, a simpler and more natural way to determine what hypotheses are missing would be to have a tool that, when fed the formula above, outputs $i \simeq j \wedge b \not\simeq c$, stating that the formula can only be true when elements $b$ and $c$ are distinct, and are inserted at the *same* position in $a$. This information permits to know immediately what additional hypotheses must be made for the formula to be unsatisfiable. In this example, there are two possible hypotheses that can be added: $i \not\simeq j$ or $b \simeq c$.

In this paper, we investigate what information should be provided to the user and how it can be obtained, by distinguishing a set of constants on which additional hypotheses are allowed to be made. These constants are called *abducible constants* or simply *abducibles*, and the problem boils down to determining what ground clauses containing only abducibles are logically entailed by the formula under consideration, since the negation of any of these clauses can be viewed as a set of additional hypotheses that make the formula unsatisfiable.

**Outline.** This paper begins by summarizing all necessary background, and then a calculus specially designed to abductive reasoning is defined. This calculus is closely related to the *superposition calculus* $\mathcal{SP}$, and we rely on completeness and termination results for $\mathcal{SP}$ to prove similar results for the new calculus. We also propose a method for generating clauses containing only abducibles, that can help a user quickly detect where an error comes from, and decide what additional hypotheses should be added to fix the faulty formula.

## 2 Preliminaries

The general framework of this paper is first-order logic with equality. Most of the presentation in this section is standard, and we refer the reader to [13] for details. Given a signature $\Sigma$ and an integer $i \geq 0$, $\Sigma^i$ stands for the set of function symbols in $\Sigma$ of arity $i$. In particular, $\Sigma^0$ denotes the set of constants in $\Sigma$. We assume the standard definitions of terms, predicates, literals and clauses, all of which are constructed over a set of variables $\mathcal{X}$. We also consider the standard definitions of positions in terms, predicates, literals or clauses; the set of positions of a term $t$ is denoted by $\mathrm{Pos}(t)$. A term, predicate, literal or clause containing no variable is *ground*. As usual, clauses are assumed to be variable-disjoint. The symbol $\simeq$ stands for unordered equality, $\bowtie$ is either $\simeq$ or $\not\simeq$. A literal $t \simeq s$ is *positive*, and a literal $t \not\simeq s$ is *negative*. If $L$ is a literal, then $L^c$ denotes the complementary literal of $L$, i.e., $(t \simeq s)^c \stackrel{\text{def}}{=} (t \not\simeq s)$ and $(t \not\simeq s)^c \stackrel{\text{def}}{=} (t \simeq s)$. A literal is *flat* if it only contains constants or variables[1], and a clause is *flat* if it only contains flat literals. The letters $l, r, s, u, v$ and $t$ denote terms, $w, x, y, z$ variables, and all other lower-case letters denote constants or function symbols.

**Definition 1** Given a ground clause $C$, we denote by $\neg C$ the following set of literals: $\neg C \stackrel{\text{def}}{=} \{L^c \mid L \in C\}$. $\Diamond$

Throughout this paper, for technical convenience, we will compare clauses modulo associativity and commutativity of the disjunction operator, but *not* modulo idempotence. For instance, the clause $f(a) \simeq f(b) \vee c \simeq d \vee a \simeq c$ will be considered as equal to $c \simeq a \vee f(b) \simeq f(a) \vee c \simeq d$, and different from $f(a) \simeq f(b) \vee c \simeq d \vee a \simeq c \vee a \simeq c$.

A *substitution* is a function mapping variables to terms. Given a substitution $\sigma$, the set of variables $x$ such that $x\sigma \neq x$ is called the *domain* of $\sigma$ and denoted by $dom(\sigma)$. If $\sigma$ is a substitution and $V$ is a set of variables, then $\sigma_{|V}$ is the substitution with domain $dom(\sigma) \cap V$, that matches $\sigma$ on this domain. As usual, a substitution can be extended into a homomorphism on terms, atoms, literals and clauses. The image of an expression $\mathcal{E}$ by a substitution $\sigma$ will be denoted by $\mathcal{E}\sigma$. If $E$ is a set of expressions, then $E\sigma$ denotes the set $\{\mathcal{E}\sigma \mid \mathcal{E} \in E\}$. The composition of two substitutions $\sigma$ and $\theta$ is denoted by $\sigma\theta$. A substitution $\sigma$ is *more general* than $\theta$ if there exists a substitution $\eta$ such that $\theta = \sigma\eta$. The substitution $\sigma$ is a *renaming* if it is injective and $\forall x \in dom(\sigma), x\sigma \in \mathcal{X}$; and it is a *unifier* of two terms $t, s$ if $t\sigma = s\sigma$. Any unifiable pair of terms $(t, s)$ has a

[1]Note that we depart from the terminology in [2, 1], where flat positive literals can contain a term of depth 1.

| | | |
|---|---|---|
| *Superposition* | $\dfrac{C \vee l[u'] \simeq r \quad D \vee u \simeq t}{(C \vee D \vee l[t] \simeq r)\sigma}$ | $(i), (ii), (iii), (iv)$ |
| *Paramodulation* | $\dfrac{C \vee l[u'] \not\simeq r \quad D \vee u \simeq t}{(C \vee D \vee l[t] \not\simeq r)\sigma}$ | $(i), (ii), (iii), (iv)$ |
| *Reflection* | $\dfrac{C \vee u' \not\simeq u}{C\sigma}$ | $(v)$ |
| *Equational Factoring* | $\dfrac{C \vee u \simeq t \vee u' \simeq t'}{(C \vee t \not\simeq t' \vee u \simeq t')\sigma}$ | $(i), (vi)$ |

where the notation $l[u']$ means that $u'$ appears as a subterm in $l$, $\sigma$ is the most general unifier (mgu) of $u$ and $u'$, $u'$ is not a variable in *Superposition* and *Paramodulation*, and the following abbreviations hold:

***(i)***: $u\sigma \not\preceq t\sigma$;

***(ii)***: $\forall L \in D : (u \simeq t)\sigma \not\preceq L\sigma$;

***(iii)***: $l[u']\sigma \not\preceq r\sigma$;

***(iv)***: $\forall L \in C : (l[u'] \bowtie r)\sigma \not\preceq L\sigma$;

***(v)***: $\forall L \in C : (u' \simeq u)\sigma \not\prec L\sigma$;

***(vi)***: $\forall L \in \{u' \simeq t'\} \cup C : (u \simeq t)\sigma \not\prec L\sigma$.

Figure 2: Inference rules of $\mathcal{SP}$: the clause below the inference line is added to the clause set containing the clauses above the inference line.

most general unifier, unique up to a renaming, and denoted by $\mathrm{mgu}(t, s)$. A substitution $\sigma$ is *ground* if $x\sigma$ is ground, for every variable $x$ in its domain.

A *simplification ordering* $\prec$ is an ordering that is stable under substitutions, monotonic and contains the subterm ordering: if $s \prec t$, then $c[s]\sigma \prec c[t]\sigma$ for any context $c$ and substitution $\sigma$, and if $s$ is a strict subterm of $t$ then $s \prec t$. A *complete simplification ordering*, or CSO, is a simplification ordering that is total on ground terms. Similarly to [7], in the sequel, we shall assume that any CSO under consideration is *good*:

**Definition 2** A CSO $\succ$ is *good* if for all ground compound terms $t$ and constants $c$, we have $t \succ c$. $\diamondsuit$

The *superposition calculus*, or $\mathcal{SP}$ (see, e.g., [13]), is a refutationally complete rewrite-based inference system for first-order logic with equality. It consists of the inference rules summarized in Fig. 2: each rule contains *premises* which are above the inference line, and generates a *conclusion*, which is below the inference line. If a clause $D$ is generated from premises $C, C'$, then we write $C, C' \vdash D$. The superposition calculus is based on a CSO on terms, which is extended to literals and clauses in a standard way (see, e.g., [3]), and we may write $\mathcal{SP}_\prec$ and $\vdash_\prec$ to specify the ordering. The set of clauses that are *deducible with $\mathcal{SP}$ from premises in $S$* is denoted by $\mathcal{I}(S)$; it consists of all clauses that are generated by the inference rules in $\mathcal{SP}$ with premises in $S$. A set of clauses $S$ is *$\mathcal{SP}$-closed* if $\mathcal{I}(S) \subseteq S$. Given a set of clauses $S$ and a clause $C$, an *$\mathcal{SP}$-derivation of $C$ from $S$* is a sequence $(C_1, \ldots, C_n)$ where $n \geq 0$, such that $C_n = C$ and for all $i \leq n$, $C_i \in S \cup \mathcal{I}(\{C_1, \ldots, C_{i-1}\})$. An *$\mathcal{SP}$-refutation of $S$* is an $\mathcal{SP}$-derivation of $\Box$ from $S$. A ground clause C is *$\prec$-redundant in $S$*, or simply redundant, if there exists a set of ground clauses $S'$ such that $S' \models C$, and for every $D \in S'$, $D$ is an instance of a clause in $S$ and $D \prec C$. A non-ground clause $C$ is *$\prec$-redundant in $S$* if all its instances are $\prec$-redundant in $S$. In particular, every strictly subsumed clause and every tautological clause is redundant. A set of clauses $S$ is *saturated* if every clause $C \notin S$ generated from premises in $S$ is redundant in $S$. A saturated set of clauses that does not contain $\Box$ is satisfiable [13]. In practice, it is necessary to use a decidable approximation of this notion of redundancy: for example, a clause is redundant if it can be reduced by some demodulation steps to either a tautology or to a subsumed clause.

In the sequel, it will be necessary to forbid the occurrence of clauses containing maximal literals of the form $x \simeq t$, where $x \not\preceq t$:

**Definition 3** A clause is *variable-eligible w.r.t.* $\prec$ if it contains a maximal literal of the form $x \simeq t$, where $x \not\preceq t$. A set of clauses is *variable-inactive* (see [1]) if no non-redundant clause generated from $S$ is variable-eligible. ◊

For technical reasons we have chosen to present a slightly relaxed version of the superposition calculus, in which the standard strict maximality conditions have been replaced by maximality conditions. For instance in Condition $i$), $u\sigma \not\preceq t\sigma$ is replaced by $u\sigma \not\prec t\sigma$: it is not forbidden for $u$ and $t$ to be distinct in *Paramodulation* and *Superposition* inferences. It is clear that the clauses generated in the case where there is an equality actually turn out to be redundant: for instance, if $u = t$ then the clause generated by the inference will be redundant w.r.t. its first premise.

## 3 A calculus for handling abducibles constants

### 3.1 Overview

As explained in the Introduction, the aim of this paper is to start with a formula $F$ and a set of axioms $A$, and generate a formula $H$ which logically entails $F$ modulo $A$, i.e., such that $H, A \models F$ (where $H \wedge A$ is satisfiable). As usual in abductive reasoning (see for instance [8]), we actually consider the contrapositive: since $H, A \models F$ is equivalent to $\neg F, A \models \neg H$, the original problem can be solved by generating logical consequences of the formula $\neg F \wedge A$. For the sake of simplicity, the formula $\neg F$ is added to the axioms which are assumed to be in clausal form, and we have the following definition:

**Definition 4** A clause $C$ is an *implicate* of a set of clauses $S$ iff $S \models C$. ◊

It is clear that after its generation, it is necessary to verify that $H$ is satisfiable modulo $A$. For instance, if $a$ is some constant, then an explanation such as $a \simeq 0 \wedge a \simeq 1$ or even $0 \simeq 1$ does not provide any information since it contradicts the axioms of Presburger arithmetic. Testing this satisfiability can be done using standard decision procedures. There are many possible candidate sets of implicates, which may be more or less informative. For instance, it is possible to take $C \in S$, but this is obviously of no use. Thus it is necessary to provide additional information in order to restrict the class of formulas that are searched for. In (propositional) abductive reasoning, this is usually done by considering clauses built on a given set of literals: the *abducible literals*. A more natural possibility in the context of this paper is to consider clauses built on a given set of ground terms. We may assume with no loss of generality that each of these terms is replaced by a constant symbol, by applying the usual flattening operation, see, e.g., [2, 7]. For example, the term $\mathrm{select}(\mathrm{store}(a, i, b), j)$ may be replaced by a new constant $d$, along with the axioms: $d \simeq \mathrm{select}(d', j) \wedge d' \simeq \mathrm{store}(a, i, b)$). We thus consider a distinguished set of constants $\mathcal{A} \subseteq \Sigma^0$, called the *set of abducible constants*, and restrict ourselves to explanations that are conjunctions of literals built on abducible constants. This is formalized with the following definition of an $\mathcal{A}$-implicate:

**Definition 5** Let $S$ be a set of clauses. A clause $C$ is an $\mathcal{A}$*-implicate* of $S$ iff every term occurring in $C$ is also in $\mathcal{A}$ and if $S \models C$. $\Diamond$

As in propositional abductive reasoning, the set $\mathcal{A}$ must be provided by the user. The elements of $\mathcal{A}$ can simply be called *abducibles*. Given a set of clauses $S$ containing both the axioms $A$ and the clauses corresponding to the conjunctive normal form of $\neg F$, we investigate how to generate the set of flat ground clauses $C$ built on $\mathcal{A}$, that are logical consequences of $S$. Since $\mathcal{SP}$ is only *refutationally* complete, this cannot be done directly using this calculus. For instance, it is clear that $f(a) \not\simeq f(b) \models a \not\simeq b$, but $a \not\simeq b$ cannot be generated from the antecedent clause. In principle, it is possible to enumerate all possible clauses $C$ built on $\mathcal{A}$ and then use the superposition calculus to check whether $S \cup \neg C$ is unsatisfiable, however, this yields a very inefficient procedure. An alternate method consists in replacing the superposition calculus by a less restrictive calculus, such as the Resolution calculus [10] together with the equality axioms. For instance in the previous case, the clause $f(a) \not\simeq f(b)$ and the substitutivity axiom $x \not\simeq y \vee f(x) \simeq f(y)$ permit to generate by the Resolution rule: $a \not\simeq b$. However, again, this calculus is not efficient, and in particular all the termination properties of the superposition calculus on many interesting subclasses of first-order logic [4, 2, 1] are lost. In this section, we provide a variant of the superposition calculus which is able to *directly* generate, from a set of clauses $S$, a set of logical consequences of $S$ that are built on a given set of constant symbols $\mathcal{A}$. The calculus is thus parameterized both by the term ordering $\prec$ and by the set of abducibles $\mathcal{A}$. We shall show that the calculus is complete, in the sense that if $S \models C$ and if $C$ is an $\mathcal{A}$-implicate of $S$, then $C$ is a logical consequence of other clauses built on $\mathcal{A}$ that are generated from $S$. We will also prove that the calculus terminates on many classes of interest in the SMT community.

We will thus consider clauses of a particular form and a slight variation of the superposition calculus in order to be able to reason on abducibles. The principle behind this calculus is similar to that of [5] for the combination of *hierarchic theories*, with the difference that in this framework, abducibles can potentially interact with other terms, whereas in the framework of [5], elements of the different theories are of different sorts. In both settings however, a same *abstraction* principle is used to delay the reasoning on the objects of interest (in this case, the abducible constants). In what follows, we formally define such an abstraction and prove some of its properties. We then formally define the calculus $\mathcal{SP}_{\mathcal{A}}$.

### 3.2 Abstraction

From now on we assume that the set of variables $\mathcal{X}$ is of the form $\mathcal{X} = \mathcal{V} \uplus \mathcal{V}_{\mathcal{A}}$. The elements in $\mathcal{V}$ are ordinary variables and the elements in $\mathcal{V}_{\mathcal{A}}$ are called *abducible variables*, and they will serve as placeholders for abducible constants in terms and clauses. In the sequel, when we mention *standard* terms, literals or clauses, we assume that all the variables they contain are in $\mathcal{V}$.

**Definition 6** An $\mathcal{A}$-*literal* is a literal of the form $t \bowtie s$, where $t, s \in \mathcal{V}_{\mathcal{A}} \cup \mathcal{A}$. An $\mathcal{A}$-*clause* is a disjunction of $\mathcal{A}$-literals. Given a clause $C$, we denote by $\Delta(C)$ the disjunction of $\mathcal{A}$-literals in $C$ and by $\overline{\Delta}(C)$ the disjunction of non-$\mathcal{A}$-literals in $C$. We denote by $\mathrm{Var}_{\mathcal{A}}(C)$ the set $\mathrm{Var}(C) \cap \mathcal{V}_{\mathcal{A}}$. ♢

A first step towards reasoning on abducibles will consist in extracting them from the terms in which they occur, and replacing them by abducible variables. Then, to ensure that such a property is preserved by inferences, every substitution mapping an abducible variable to anything other than an abducible variable will be discarded. More formally:

**Definition 7** A term is *abstracted* if it contains no abducible constant. A literal $t \bowtie s$ is *abstracted* if $t$ and $s$ are both abstracted. A clause is *abstracted* if all non-abstracted literals in $C$ are $\mathcal{A}$-literals. ♢

If $t$ is an abstracted term, then not every instance of $t$ is also abstracted. We define a condition on substitutions that guarantees such a stability result.

**Definition 8** A substitution $\sigma$ is $\mathcal{A}$-*compliant* if for all $x \in dom(\sigma)$, $x\sigma$ is abstracted, and for all $x \in dom(\sigma) \cap \mathcal{V}_{\mathcal{A}}$, $x\sigma \in \mathcal{V}_{\mathcal{A}}$. Two abstracted terms are $\mathcal{A}$-*unifiable* if they are unifiable and admit an $\mathcal{A}$-compliant mgu. ♢

**Proposition 9** *If $\sigma$ and $\mu$ are $\mathcal{A}$-compliant, then so is $\sigma\mu$. If $\sigma$ is $\mathcal{A}$-compliant and $t$ is abstracted, then so is $t\sigma$.*

It will be possible to define a calculus that generates abstracted clauses from abstracted premises thanks to the following property:

**Proposition 10** *If the abstracted terms $t, s$ are unifiable and admit an mgu $\mu$ such that for all $x \in \mathcal{V}_{\mathcal{A}}$, $x\mu \in \mathcal{V} \cup \mathcal{V}_{\mathcal{A}}$, then $t, s$ are $\mathcal{A}$-unifiable.*

Proof. If $x \in \mathcal{V}_{\mathcal{A}}$, $y \in \mathcal{V}$ and $x\mu = y$, then $\mu' = \mu\{y \mapsto x\}$ is also an mgu of $t, s$ since it is a renaming of $\mu$, and $dom(\mu') \cap \mathcal{V}_{\mathcal{A}} = dom(\mu) \cap (\mathcal{V}_{\mathcal{A}} \setminus \{x\})$. By repeating this operation on all variables $x, y$ such that $x \in \mathcal{V}_{\mathcal{A}}$, $y \in \mathcal{V}$ and $x\mu = y$, we eventually obtain an mgu $\mu''$ of $t, s$ that, by construction, is $\mathcal{A}$-compliant. ■

In the sequel, every time abstracted terms are $\mathcal{A}$-unifiable, we will assume the corresponding mgu is $\mathcal{A}$-compliant. The following definition shows how abstracted terms can be transformed into standard ones by replacing all variables in $\mathcal{V}_\mathcal{A}$ by an arbitrary element in $\mathcal{A}$.

**Definition 11** Let $<_\mathcal{A}$ be a total ordering on $\mathcal{A}$ and $a_0$ denote the smallest abducible in $\mathcal{A}$. Given a term $t$, we denote by $t_{\downarrow\mathcal{A}}$ the term obtained by replacing every abducible occurring in $t$ by $a_0$. The term $t$ is *$\mathcal{A}$-reduced* if $t_{\downarrow\mathcal{A}} = t$. The previous notation and this definition extend to literals, clauses and sets of clauses. ♢

**Example 12** Let $C = f(b,c) \simeq g(d) \vee x \not\simeq b \vee f(a,b) \not\simeq f(c,d)$, where $\mathcal{A} = \{a,b,c\}$ and $a \prec b \prec c$. Then $C_{\downarrow\mathcal{A}} = f(a,a) \simeq g(d) \vee x \not\simeq a \vee f(a,a) \not\simeq f(a,d)$, and this clause is an $\mathcal{A}$-reduced clause.

### $\mathcal{V}_\mathcal{A}$-stability

It is clear that if all abducibles are abstracted away from a standard clause, then the resulting abstracted clause is not equivalent to the former one. However, equivalence can be regained by adding so-called *$\mathcal{V}_\mathcal{A}$-constraint literals* to the resulting abstracted clause.

**Definition 13** A *$\mathcal{V}_\mathcal{A}$-constraint literal* is a literal of the form $x \not\simeq a$, where $x \in \mathcal{V}_\mathcal{A}$ and $a \in \mathcal{A}$. For all clauses $C$, we denote by $\Gamma(C)$ the disjunction of $\mathcal{V}_\mathcal{A}$-constraint literals in $C$. A *$\mathcal{V}_\mathcal{A}$-constraint clause* is a disjunction of $\mathcal{V}_\mathcal{A}$-constraint literals. Given a $\mathcal{V}_\mathcal{A}$-constraint clause $A = \bigvee_{i=1}^{k} x_i \not\simeq a_i$, the *substitution associated to $A$* is denoted by $\nu_A$ and defined as follows: $dom(\nu_A) = \{x_1, \ldots, x_k\}$, and for all $x \in dom(\nu_A)$, $x\nu_A = \min_{<_\mathcal{A}} \{a_i \mid x_i = x\}$.

For readability, if $B$ is a clause then we will write $\nu_B$ instead of $\nu_{\Gamma(B)}$. If $S$ is a set of abstracted clauses, then $S_\nu$ is the set $S_\nu = \{C\nu_C \mid C \in S\}$. ♢

**Example 14** Assume $\mathcal{A} = \{a,b,c\}$, where $a <_\mathcal{A} b <_\mathcal{A} c$, and let $A = x \not\simeq a \vee x \not\simeq c \vee y \not\simeq b \vee z \not\simeq a \vee y \not\simeq c$. Then $\nu_A = \{x \mapsto a, y \mapsto b, z \mapsto a\}$.

Note that by definition, $C \equiv C\nu_C$ and $S \equiv S_\nu$. As mentioned earlier, abducible variables are meant to be placeholders for abducible constants. In general, it will be necessary to keep some information permitting to know what abducible constants an abducible variable could be replaced by. Such a requirement is satisfied by imposing that every abducible variable occurs in at least one $\mathcal{V}_\mathcal{A}$-constraint literal, which intuitively specifies its value.

**Definition 15** A clause $C$ is *$\mathcal{V}_\mathcal{A}$-stable* if $\mathrm{Var}_\mathcal{A}(C) \subseteq \mathrm{Var}_\mathcal{A}(\Gamma(C))$. A set of clauses is *$\mathcal{V}_\mathcal{A}$-stable* if every clause it contains is $\mathcal{V}_\mathcal{A}$-stable. ♢

Note that if $C$ is such that $\mathrm{Var}(C) \subseteq \mathcal{V}_\mathcal{A}$, then $C$ is $\mathcal{V}_\mathcal{A}$-stable if and only if $C\nu_C$ is ground. For example, if $C$ is a $\mathcal{V}_\mathcal{A}$-stable $\mathcal{A}$-clause, then $C\nu_C$ is ground.

**Lemma 16** *Let $C$ be an $\mathcal{A}$-clause that is $\mathcal{V}_\mathcal{A}$-stable, $\mu$ be a substitution whose codomain is contained in $\mathrm{Var}_\mathcal{A}(C)$, and let $I$ be an interpretation such that:*

*1. for all $x \in \mathrm{Var}_{\mathcal{A}}(C)$, $I \models x\nu_C \simeq x\mu\nu_C$,*

*2. $I \models \neg(C\nu_C)$.*

*Let $D = C\mu$, then for all $x \in \mathrm{Var}_{\mathcal{A}}(D)$, $I \models x\nu_C \simeq x\nu_D$. Thus, in particular, $I \models \neg(D\nu_D)$.*

PROOF. Let $x \in \mathrm{Var}_{\mathcal{A}}(D)$, $\nu = \nu_C$ and $\nu' = \nu_D$. By hypothesis $x$ is in $\mathrm{Var}_{\mathcal{A}}(C)$, and for all literals $x \not\simeq b$ occurring in $C$, $I \models x\nu \simeq b$. Also, $I \models x\nu \simeq x\mu\nu$, thus $I \models x\mu\nu \simeq b$. Assume $x\nu' = c$ for some abducible constant $c \in \mathcal{A}$. By definition, $D$ must contain a literal of the form $x \not\simeq c$, thus $C$ must contain a literal $y \not\simeq c$, where $y\mu = x$. Since $I \models \neg(C\nu)$, necessarily $I \models y\nu \simeq c$, whence $I \models y\mu\nu \simeq c$ i.e. $I \models x\nu \simeq c$. Thus $I \models x\nu \simeq x\nu'$ and $I \models \neg(D\nu')$. ■

Given a set of standard clauses, it is easy to construct an equivalent set of abstracted and $\mathcal{V}_{\mathcal{A}}$-stable clauses. It suffices to replace every abducible $a$ occurring in a non-$\mathcal{A}$-literal by a fresh variable $x \in \mathcal{V}_{\mathcal{A}}$, and to add the literal $x \not\simeq a$ to the clause. For instance, if $\mathcal{A} = \{a, b\}$ then the clause $a \simeq b \vee a \simeq c \vee f(b, d, x) \not\simeq g(b, y)$ is replaced by $x_1 \not\simeq a \vee x_2 \not\simeq b \vee x_3 \not\simeq b \vee a \simeq b \vee x_1 \simeq c \vee f(x_2, d, x) \not\simeq g(x_3, y)$. Note that if $C$ is already an $\mathcal{A}$-clause, then the abstracted form of $C$ is $C$ itself.

### 3.3 Definition of the calculus.

We introduce a calculus for generating $\mathcal{A}$-implicates. It is a modified version of the superposition calculus, and consists of inference rules that are meant to be applied to abstracted clauses. In particular, it is based on orderings that are suitable for abstracted terms, literals and clauses: the order between two terms $t$ and $s$ should not depend on the abducible constants occurring in $t$ and $s$, and maximal terms and literals in abstracted clauses should be related to maximal terms and literals in standard clauses, in a sense that will be made precise later. We thus define particular orderings for standard clauses, from which we define suitable orderings for abstracted clauses.

**Definition 17** We consider a good complete simplification ordering $\prec$ such that:

1. for all $a, b \in \mathcal{A}$, $a \prec b$ if and only if $a <_{\mathcal{A}} b$;

2. for all $a \in \mathcal{A}$ and for all non-variable terms $t \notin \mathcal{A}$, $a \prec t$;

3. for all ground terms $t, s$ not in $\mathcal{A}$, if $t \prec s$ then $t_{\downarrow\mathcal{A}} \preceq s_{\downarrow\mathcal{A}}$, and if $t_{\downarrow\mathcal{A}} \prec s_{\downarrow\mathcal{A}}$ then $t \prec s$.

We let $\gamma_0$ denote the ground substitution of domain[2] $\mathcal{V}_{\mathcal{A}}$ such that for all $x \in \mathcal{V}_{\mathcal{A}}$, $x\gamma_0 = a_0$. Given abstracted terms $t, s$, we define $\prec_{\mathcal{A}}$ as follows: $t \prec_{\mathcal{A}} s$ iff $t\gamma_0 \prec s\gamma_0$. This definition extends to literals and clauses in a standard way. A term is $\mathcal{A}$-*maximal* if it is maximal for $\prec_{\mathcal{A}}$; this definition also extends to literals and clauses. ◊

[2]Note that the domain of $\gamma_0$ is infinite. This does not cause any technical problem and allows the expression of several properties in a concise way.

It is not difficult to construct a good CSO that satisfies the requirements of Definition 17, one such construction goes as follows: consider any good (decidable) CSO $\prec_0$ that is defined on the set $T$ of ground $\mathcal{A}$-reduced terms, and such that $a_0 \prec_0 b \prec_0 t$ for all constants $b \neq a_0$ and compound terms $t$ in $T$. Let $T'$ denote the set of all ground terms constructed over the signature, and for $t \in T'$, define $[t]_{\mathcal{A}} = \left\{t' \in T' \mid t'_{\downarrow\mathcal{A}} = t_{\downarrow\mathcal{A}}\right\}$. Then inductively define the order $\prec$ on $T'$ as follows:

- for all $t, s \in T'$, if $t_{\downarrow\mathcal{A}} \prec_0 s_{\downarrow\mathcal{A}}$ then $t' \prec s'$ for all $t' \in [t]_{\mathcal{A}}$ and $s' \in [s]_{\mathcal{A}}$;
- for all $t \in T'$ and for all $s, s' \in [t]_{\mathcal{A}}$,
  - if $t_{\downarrow\mathcal{A}} = a_0$ then $s \prec s'$ iff[3] $s <_{\mathcal{A}} s'$,
  - otherwise, $t = f(t_1, \ldots, t_n)$, in which case $s = f(s_1, \ldots, s_n)$ and $s' = f(s'_1, \ldots, s'_1)$, and $s \prec s'$ iff $\langle s_1, \ldots, s_n\rangle \prec^{lex} \langle s'_1, \ldots, s'_n\rangle$, where $\prec^{lex}$ denotes the lexicographic extension of $\prec$.

The ordering $\prec$ can then be straightforwardly extended to non-ground terms in such a way that Condition 2 of Definition 17 is satisfied. It is simple to verify that $\prec$ is an ordering which is total on ground terms and stable under substitutions. It satisfies the subterm property because if $s = t|_p$, $s' = s_{\downarrow\mathcal{A}}$ and $t' = t_{\downarrow\mathcal{A}}$, then $s' = t'|_p$. Thus $s' \prec_0 t'$ and $s \prec t$ by construction. It is also stable under operations: indeed, assume $s \prec s'$ and consider the terms

$$t \;=\; f(t_1, \ldots, t_{i-1}, s, t_{i+1}, \ldots, t_n) \text{ and } t' \;=\; f(t'_1, \ldots, t'_{i-1}, s', t'_{i+1}, \ldots, t'_n).$$

If $s_{\downarrow\mathcal{A}} \prec_0 s'_{\downarrow\mathcal{A}}$ then $t_{\downarrow\mathcal{A}} \prec_0 t'_{\downarrow\mathcal{A}}$ because $\prec_0$ is a CSO, and by construction $t \prec t'$. Otherwise $s_{\downarrow\mathcal{A}} = s'_{\downarrow\mathcal{A}}$ and

$$\langle t_1, \ldots, t_{i-1}, s, t_{i+1}, \ldots, t_n\rangle \;\prec^{lex}\; \langle t_1, \ldots, t_{i-1}, s', t_{i+1}, \ldots, t_n\rangle,$$

hence again, $t \prec t'$. Therefore, $\prec$ is a CSO, and by construction, this CSO is good. Note that the order $\prec$ used is not necessarily decidable, but this does not matter since this order will be used only for theoretical purposes, it is not intended to be used in a concrete proof procedure.

The following propositions are entailed by the properties of the ordering under consideration:

**Proposition 18**

1. *If $C$ is a non-variable-eligible clause containing a $\prec$-maximal literal with a $\prec$-maximal term in $\mathcal{A}$, then $C$ is an $\mathcal{A}$-clause.*
2. *Let $C$ be an abstracted clause that is not an $\mathcal{A}$-clause. If $L\nu_C$ is $\prec$-maximal in $C\nu_C$, then $L$ is $\mathcal{A}$-maximal in $C$. Furthermore, if $L\nu_C = (t \bowtie s)\nu_C$ and $t\nu_C$ is $\prec$-maximal in $L\nu_C$, then $t$ is $\mathcal{A}$-maximal in $L$.*

[3]Note that in this case, both $s$ and $s'$ must be in $\mathcal{A}$.

**Definition 19** We denote by $\mathcal{SP}_{\mathcal{A}}$ the calculus such that for all clause sets $S$, we have $S \vdash^{\mathcal{A}} D$ if $S \vdash_{\prec_{\mathcal{A}}} D$ and the mgu involved in the $\mathcal{SP}$-inference is $\mathcal{A}$-compliant. $\Diamond$

By construction, $\mathcal{SP}$ and $\mathcal{SP}_{\mathcal{A}}$ coincide on ground $\mathcal{A}$-clauses.

**Redundancy Elimination for Abstracted Clauses**

We define a particular notion of redundancy for abstracted clauses, that is related to redundancy for standard clauses. The main difference with the standard definition is that the redundancy test is performed modulo the substitution $\nu_C$ that replaces the abstracted variables in $C$ by the abducibles they denote.

**Definition 20** Consider a set of abstracted clauses $S$ and an abstracted clause $C$ such that $\text{Var}(C) \subseteq \mathcal{V}_{\mathcal{A}}$. The clause $C$ is *$\mathcal{A}$-redundant in $S$* if one of the following condition holds:

- $C$ is an $\mathcal{A}$-clause, $\nu_C \neq \text{id}$ and $C\nu_C$ occurs in $S$ or is $\mathcal{A}$-redundant in $S$,
- there exists a set of ground clauses $S'$ such that $S' \models C$, every $D \in S'$ is an instance of a clause in $S_\nu$ and $D \prec C\nu_C$.

If $C$ is an abstracted clause such that $\text{Var}(C) \not\subseteq \mathcal{V}_{\mathcal{A}}$, then $C$ is *$\mathcal{A}$-redundant in $S$* if for all ground substitutions $\sigma$ with a domain in $\mathcal{V}$, $C\sigma$ is $\mathcal{A}$-redundant in $S$. The set $S$ is *$\mathcal{A}$-saturated* if every clause $C \notin S$ generated by an $\mathcal{SP}_{\mathcal{A}}$-inference with premises in $S$ is $\mathcal{A}$-redundant in $S$. $\Diamond$

This notion of redundancy permits to add the standard contraction rules of the superposition calculus to $\mathcal{SP}_{\mathcal{A}}$ (subsumption, simplification, elimination of tautologies, etc). The following contraction inference rule is also added to $\mathcal{SP}_{\mathcal{A}}$:

$$\mathcal{A}\text{-reduction}: \quad \frac{C}{C\nu_C} \quad \text{if } C \text{ is an } \mathcal{A}\text{-clause and } \nu_C \neq \text{id}.$$

After any application of the $\mathcal{A}$-reduction rule, the premise becomes $\mathcal{A}$-redundant and can be deleted.

**Theorem 21** *If $S$ is a variable-inactive set of abstracted clauses that are $\mathcal{V}_{\mathcal{A}}$-stable, then every non-redundant clause generated from $S$ by $\mathcal{SP}_{\mathcal{A}}$ is abstracted and $\mathcal{V}_{\mathcal{A}}$-stable. Also, if one of the premises of a binary $\mathcal{SP}_{\mathcal{A}}$-inference is an $\mathcal{A}$-clause, then the other premise is also an $\mathcal{A}$-clause.*

Proof. The first property is a consequence of the fact that if $C$ is abstracted and $\mathcal{V}_{\mathcal{A}}$-stable, and if $\sigma$ is an $\mathcal{A}$-compliant substitution, then $C\sigma$ is also an abstracted and $\mathcal{V}_{\mathcal{A}}$-stable clause. Since the $\mathcal{A}$-maximal term in a positive $\mathcal{A}$-maximal literal of a premise cannot be a variable, a non-abducible term cannot be replaced by an abducible constant, and thus it is straightforward to verify that the clause generated by $\mathcal{SP}$ is abstracted and $\mathcal{V}_{\mathcal{A}}$-stable. The second point is a direct consequence of Proposition 18 (1). ∎

In what follows, we will prove completeness and termination results for $\mathcal{SP}_\mathcal{A}$. The completeness result guarantees that $\mathcal{SP}_\mathcal{A}$ generates the required information about existing abducibles for any abstracted set of clauses, while the termination result relies on termination results for $\mathcal{SP}$, and will be used to verify without any additional effort that our technique can be used as a decision procedure for reasoning about abducibles in SMT problems with several theories of interest.

## 4 Completeness of the calculus

This section is devoted to the proof that if $S$ is an unsatisfiable set of abstracted clauses that is $\mathcal{A}$-saturated, then $\Box \in S$. Note that this result does *not* follow from the refutational completeness of the superposition calculus: indeed, the ordering $\prec_\mathcal{A}$ is not a simplification ordering (it is not stable by substitution), and all inferences in which non-$\mathcal{A}$-compliant unifiers are involved are ignored. However, the proof is based on the refutational completeness of $\mathcal{SP}$, and requires determining relationships between $\mathcal{SP}$-inferences and $\mathcal{SP}_\mathcal{A}$-inferences. The following properties relate mgus of abstracted terms to mgus of corresponding standard terms.

**Lemma 22** *Let $t, s$ be abstracted terms, $\delta$ be a substitution with a domain in $\mathcal{V}_\mathcal{A}$, such that for all $x \in \mathrm{Var}_\mathcal{A}(t) \cup \mathrm{Var}_\mathcal{A}(s)$, $x\delta \in \mathcal{A}$, and consider the standard terms $t' = t\delta$ and $s' = s\delta$. If $t'$ and $s'$ are unifiable, then $t, s$ are $\mathcal{A}$-unifiable, and if $\mu$ is an mgu of $t, s$ then $\delta\mu\delta = \mu\delta$ and $(\mu\delta)_{|\mathcal{V}}$ is an mgu of $t', s'$.*

Proof. Let $\gamma'$ be an mgu of $t'$ and $s'$, the result is proved by induction on the size of $t'\gamma'$. If one of $t'$ or $s'$ is in $\mathcal{A} \cup \mathcal{V}$, then the result is not difficult to verify. For instance, if $t' \in \mathcal{A}$ and $s' \in \mathcal{V}$, then $t \in \mathcal{V}_\mathcal{A}$, $s = s'$ and $\delta$ contains the mapping $t \mapsto t'$. In this case, $t$ and $s$ are indeed $\mathcal{A}$-unifiable with $\mu = \{s' \mapsto t\}$, and $\mu' = (\mu\delta)_{|\mathcal{V}} = \{s' \mapsto t'\}$ is an mgu of $t', s'$. Similar reasonings are carried out in the other cases.

Now assume that $t = f(t_1, \ldots, t_n)$ and $s = f(s_1, \ldots, s_n)$, so that $t' = f(t'_1, \ldots, t'_n)$ and $s' = f(s'_1, \ldots, s'_n)$. We let $\pi'_0 \stackrel{\text{def}}{=} \mathrm{id}$, and for $i = 1, \ldots, n$, $\mu'_i$ denotes the mgu of $t'_i\pi'_{i-1}$ and $s'_i\pi'_{i-1}$, and we let $\pi'_i \stackrel{\text{def}}{=} \pi'_{i-1}\mu'_i$. Since $t'$ and $s'$ are unifiable, for all $i = 1, \ldots, n$, $t'_i\pi'_{i-1}$ and $s'_i\pi'_{i-1}$ are unifiable. Furthermore, $\mu' \stackrel{\text{def}}{=} \theta'_n$ is also an mgu of $t', s'$.

Let $\pi_0 = \mathrm{id}$ and for all $i = 1, \ldots, n$, let $\mu_i$ denote the mgu of $t_i\pi_{i-1}$ and $s_i\pi_{i-1}$, and let $\pi_i = \pi_{i-1}\mu_i$. We show by induction on $i$ that $t_i\pi_{i-1}$ and $s_i\pi_{i-1}$ are $\mathcal{A}$-unifiable, that $\mu_i$ verifies $\delta\mu_i\delta = \mu_i\delta$ and $\mu'_i = (\mu_i\delta)_{|\mathcal{V}}$ and that $\pi_i$ verifies $\pi'_i = (\pi_i\delta)_{|\mathcal{V}}$. This will permit to conclude that $t$ and $s$ are $\mathcal{A}$-unifiable with mgu $\mu \stackrel{\text{def}}{=} \pi_n$ which verifies $\delta\mu\delta = \mu\delta$, and that $\mu' = (\mu\delta)_{|\mathcal{V}}$.

Assume this result holds for $i - 1$, then it is straightforward to check that $\delta\pi_{i-1}\delta = \pi_{i-1}\delta$. Consider the terms $t_i\pi_{i-1}$ and $s_i\pi_{i-1}$. By hypothesis, $\pi'_{i-1} = (\pi_{i-1}\delta)_{|\mathcal{V}}$, thus, $t'_i\pi'_{i-1} = t_i\delta(\pi_{i-1}\delta)_{|\mathcal{V}}$. Since $t_i\delta$ contains no variable in $\mathcal{V}_\mathcal{A}$, we have $t_i\delta(\pi_{i-1}\delta)_{|\mathcal{V}} = t_i\delta\pi_{i-1}\delta = t_i\pi_{i-1}\delta$. Since the size of $t'_i\pi'_{i-1}$ is strictly less than that of $t'\gamma'$, we may apply the induction hypothesis to conclude that $t_i\pi_{i-1}$ and $s_i\pi_{i-1}$ are $\mathcal{A}$-unifiable with mgu $\mu_i$ such that $\delta\mu_i\delta = \mu_i\delta$ and $\mu'_i = (\mu_i\delta)_{|\mathcal{V}}$. Therefore $\pi_i$ is well-defined and since $\pi_{i-1}\delta$ maps every variable in its domain to terms containing no variables in $\mathcal{V}_\mathcal{A}$, we have

$$\pi'_i \;=\; \pi'_{i-1}\mu'_i \;=\; (\pi_{i-1}\delta)_{|\mathcal{V}}(\mu_i\delta)_{|\mathcal{V}} \;=\; (\pi_{i-1}\delta\mu_i\delta)_{|\mathcal{V}} \;=\; (\pi_i\delta)_{|\mathcal{V}}.$$

This proves the result. ■

The following definition and properties will be used to express another useful relationship between $\mathcal{A}$-variant terms and $\mathcal{A}$-unifiable terms with slightly more general hypotheses than those of Lemma 22.

**Definition 23** Given the abstracted terms $t, s$, we denote by $\sim_{\langle t,s\rangle}$ the smallest equivalence relation such that for all variables $x$ and terms $u$, $x \sim_{\langle t,s\rangle} u$ if there exists a position $p \in \mathrm{Pos}(t) \cap \mathrm{Pos}(s)$ such that $\{x, u\} = \{t|_p, s|_p\}$. We denote by $[u]_{\langle t,s\rangle}$ (or simply by $[u]$) the equivalence class of $u$ for $\sim_{\langle t,s\rangle}$. ♢

Intuitively, if $t$ and $s$ are unifiable, then the image by their mgu of every variable $x$ occurring in $t$ or $s$ must be equal to instances of every term occurring in $[x]$. In particular, the following property holds for $\mathcal{A}$-unifiable terms:

**Proposition 24** *If $t, s$ are $\mathcal{A}$-unifiable with mgu $\mu$, then for all $x \in \mathcal{V}_\mathcal{A}$, $[x]_{\langle t,s\rangle} \subseteq \mathcal{V} \cup \mathcal{V}_\mathcal{A}$ and $x\mu \in [x]_{\langle t,s\rangle}$.*

**Lemma 25** *Let $t, s$ be abstracted terms, $\nu, \nu'$ be substitutions with identical domains contained in $\mathcal{V}_\mathcal{A}$ and with codomains in $\mathcal{A}$, and let $I$ be an interpretation such that for all $x \in \mathrm{Var}_\mathcal{A}(t) \cup \mathrm{Var}_\mathcal{A}(s)$, $I \models x\nu \simeq x\nu'$. If $t\nu'$ and $s\nu'$ are unifiable, then $t$ and $s$ are $\mathcal{A}$-unifiable, and if $\mu$ is their mgu then $I \models x\nu \simeq x\mu\nu$.*

Proof. Let $\mu'$ be the mgu of $t\nu', s\nu'$, and let $\gamma' = \nu \cup \mu'$. We prove that for all $x \in \mathcal{V}_\mathcal{A}$ and for all $y \in [x]_{\langle t,s\rangle}$, $I \models x\gamma' \simeq y\gamma'$.

Let $p$ be a position and $y, y'$ be variables in $[x]_{\langle t,s\rangle}$ such that $\{y, y'\} = \{t|_p, s|_p\}$. By hypothesis $I \models \{t|_p\nu \simeq t|_p\nu',\ s|_p\nu \simeq s|_p\nu'\}$, thus, $I \models \{t|_p\gamma' \simeq t|_p\nu'\mu',\ s|_p\gamma' \simeq s|_p\nu'\mu'\}$. But since $t\nu'\mu' = s\nu'\mu'$, necessarily $I \models t|_p\gamma' \simeq s|_p\gamma'$, i.e., $I \models y\gamma' \simeq y'\gamma'$. By transitivity, we deduce that for all $y \in [x]_{\langle t,s\rangle}$, $I \models x\gamma' \simeq y\gamma'$. By Proposition 24 $x\mu \in [x]_{\langle t,s\rangle}$, thus $I \models x\gamma' \simeq x\mu\gamma'$. Since $x$ and $x\mu$ are in $\mathcal{V}_\mathcal{A}$, we have the result. ■

We now prove that if $S$ is a set of abstracted clauses that does not contain the empty clause and is:

- $\mathcal{V}_\mathcal{A}$-stable,
- with no variable-eligible clause,
- $\mathcal{A}$-saturated,

then $S$ is satisfiable. We will show that $S$ is satisfiable by constructing a set of standard clauses whose satisfiability will entail that of $S$. The set we construct will be saturated under $\mathcal{SP}_{\prec}$-inferences, and it will not contain the empty clause; we will conclude that it must be satisfiable, and hence that so must $S$.

Let $T$ be the set of $\mathcal{A}$-clauses in $S$. Since $S$ is $\mathcal{V}_{\mathcal{A}}$-stable and $\mathcal{A}$-saturated by hypothesis, $T$ can only contain ground $\mathcal{A}$-clauses, because if a non-ground clause occurs in $T$ then $\mathcal{A}$-reduction applies. Since $\mathcal{SP}$ and $\mathcal{SP}_{\mathcal{A}}$ coincide on ground $\mathcal{A}$-clauses, $T$ must also be saturated under $\mathcal{SP}_{\prec}$-inferences and cannot contain $\Box$; this set is therefore satisfiable. We consider a fixed model of $T$.

**Definition 26** We define the ground set

$$U_I \ = \ \left\{a \simeq b \mid a, b \in \mathcal{A}, a^I = b^I\right\} \cup \left\{a \not\simeq b \mid a, b \in \mathcal{A}, a^I \neq b^I\right\}. \qquad \Diamond$$

The set $U_I$ will be used to discard all $\mathcal{A}$-clauses in $S$ in the upcoming proof. Note that by construction, $U_I$ is saturated.

**Definition 27** We inductively define the notion of an *$I$-reduction*:

- For all $a \in \mathcal{A}$, $a_{\|I} = \min_{\prec} \left\{b \in \mathcal{A} \mid b^I = a^I\right\}$.
- $f(t_1, \ldots, t_n)_{\|I} = f(t_{1\|I}, \ldots, t_{n\|I})$.

This definition extends to standard literals and clauses. $\Diamond$

The $I$-reduction procedure is used to define a set whose satisfiability entails that of $S$, and that turns out to be saturated:

**Definition 28** Let $S_I \ = \ U_I \cup \left\{\overline{\Delta}(C_{\|I}) \mid C \in S_\nu \wedge U_I \models \neg\Delta(C)\right\}$.

By construction, every $\mathcal{A}$-clause in $S_I$ must be in $U_I$ and $\Box \notin S_I$.

**Proposition 29** *If $S_I$ is satisfiable then so is $S_\nu$, and therefore so is $S$.*

Proof. Since $S_I$ contains $U_I$, necessarily $S_I \models S_\nu \models S$. ■

**Lemma 30** *$S_I$ is saturated for $\mathcal{SP}_{\prec}$.*

Proof. We prove the result by considering a clause generated by a superposition inference with premises in $S_I$, the proof in the other cases is similar. Thus assume that $C_1', C_2' \vdash_{\prec} C'$, where $C_1', C_2'$ are in $S_I$. If both $C_1'$ and $C_2'$ are $\mathcal{A}$-clauses (i.e., if they are both in $U_I$), then it is clear that $C'$ must be subsumed by a clause in $U_I$. By construction the abducible constants that occur in $S_I$ are all minimal and cannot be replaced, thus there can be no inference with one premise in $U_I$ and the other not in $U_I$. We therefore assume neither $C'_1$ nor $C'_2$ is in $U_I$. The considered clauses are of the following forms:

$$\begin{array}{rcl} C'_1 & = & u' \simeq v' \vee E'_1, \\ C'_2 & = & t'[w'] \bowtie s' \vee E'_2, \\ C' & = & (t'[v'] \bowtie s' \vee E'_1 \vee E'_2)\mu', \end{array}$$

where $\mu'$ is the mgu of $u'$ and $w'$. Since no clause in $S$ is variable-eligible by hypothesis, the maximal literals in $C'_1$ and $C'_2$ must contain symbols in $\Sigma \setminus \mathcal{A}$. By definition of $S_I$, there are clauses $C_1, C_2$ in $S$ such that:

- $U_I \models \neg\Delta(C_1\nu_{C_1})$ and $U_I \models \neg\Delta(C_2\nu_{C_2})$,
- $C'_1 = \overline{\Delta}(C_1\nu_{C_1\|I})$ and $C'_2 = \overline{\Delta}(C_2\nu_{C_2\|I})$.

Necessarily, $C_1$ and $C_2$ are of the form

$$\begin{array}{rcl} C_1 & = & u \simeq v \vee E_1, \\ C_2 & = & t[w] \bowtie s \vee E_2, \end{array}$$

and by Proposition 18 (2), $u$ is $\mathcal{A}$-maximal in $u \simeq v$ which is $\mathcal{A}$-maximal in $C_1$, and $t[w]$ is $\mathcal{A}$-maximal in $t[w] \bowtie s$ which is $\mathcal{A}$-maximal in $C_2$. Furthermore, these literals cannot be $\mathcal{A}$-literals. Therefore, $C_1, C_2 \vdash^{\mathcal{A}} C$, where $C$ is of the form

$$C \;=\; (t[v] \bowtie s \vee E_1 \vee E_2)\mu,$$

and $\mu$ is the mgu of $u, w$. We prove that $U_I \cup \{C\nu_C\} \models C'$.

It is necessary to distinguish two cases, depending on whether $(t[v] \bowtie s)\mu$ is an $\mathcal{A}$-literal or not. We let $L = (t[v] \simeq s)\mu$ if $(t[v] \simeq s)\mu$ is an $\mathcal{A}$-literal and $L = \Box$ otherwise. We also let $D = (E_1 \vee E_2)\mu$, so that the following equalities hold:

$$C \;=\; L \vee D, \tag{3}$$
$$\Delta(C) \;=\; L \vee \Delta(D), \tag{4}$$
$$\overline{\Delta}(C) \;=\; \overline{\Delta}(D), \tag{5}$$
$$\Delta(D) \;=\; \Delta((C_1 \vee C_2)\mu). \tag{6}$$

By construction, $L$ cannot be a $\mathcal{V}_{\mathcal{A}}$-constraint literal, thus $\nu_D = \nu_C$, and therefore, $C\nu_C = (L \vee D)\nu_D$. For the sake of readability, we define $\nu = \nu_{C_1} \cup \nu_{C_2}$, hence $\Delta(C_1\nu_{C_1}) \vee \Delta(C_2\nu_{C_2}) = \Delta((C_1 \vee C_2)\nu)$. We also let $\delta$ be the substitution defined as follows: for all $x \in \mathrm{Var}_{\mathcal{A}}(C_1 \vee C_2)$, $x\delta = (x\nu)_{\|I}$. Thus $C_1\delta = C'_1$ and $C_2\delta = C'_2$, and for all $x \in \mathrm{Var}_{\mathcal{A}}(C_1 \vee C_2)$, $U_I \models x\nu \simeq x\delta$. We first prove that $U_I \models \neg\Delta(D\nu_D)$.

- Since the terms $u' = u\delta$ and $w' = w\delta$ are unifiable and since $\mu$ is the mgu of $u$ and $w$, Lemma 25 proves that $U_I \models x\nu \simeq x\mu\nu$ for all $x \in \mathrm{Var}_{\mathcal{A}}(C_1 \vee C_2)$, thus Condition 1 of Lemma 16 holds for the clause $\Delta(C_1 \vee C_2)$.

- By hypothesis, $U_I \models \neg\Delta(C_1\nu_{C_1})$ and $U_I \models \neg\Delta(C_2\nu_{C_2})$; in other words, $U_I \models \neg\Delta((C_1 \vee C_2)\nu)$, and therefore, by using the equality $\Delta((C_1 \vee C_2)\nu) = \Delta(C_1 \vee C_2)\nu$, we deduce that $U_I \models \neg\Delta(C_1 \vee C_2)\nu$. Therefore, Condition 2 of Lemma 16 also holds, and using Equation 6 above, we deduce that for all $x \in \mathrm{Var}_{\mathcal{A}}(D)$ we have $U_I \models x\nu \simeq x\nu_D$ and that $U_I \models \neg\Delta(D\nu_D)$.

We have just proved that for all $x \in \mathrm{Var}_{\mathcal{A}}(C_1 \vee C_2)$, $U_I \models x\nu \simeq x\nu_D$. Since $U_I \models x\nu \simeq x\delta$, necessarily $U_I \models x\nu_D \simeq x\delta$. But every variable in $\mathrm{Var}_{\mathcal{A}}(L \vee D)$ is also in $\mathrm{Var}_{\mathcal{A}}(C_1 \vee C_2)$, therefore $U_I \cup \{\overline{\Delta}(D\nu_D)\} \models \overline{\Delta}(D\delta)$ and $U_I \cup \{L\nu_D\} \models L\delta$. In the case where $L \neq \Box$, by Lemma 22,

$$L\delta \;=\; (t[v] \simeq s)\mu\delta \;=\; (t[v] \simeq s)\delta\mu\delta \;=\; (t'[v'] \simeq s')\mu',$$

and similarly, $\overline{\Delta}(D\delta) = (E'_1 \vee E'_2)\mu'$. Therefore, it is always the case that $(L \vee \overline{\Delta}(D))\delta = C'$. Since $\mathrm{Var}(C) \cap \mathcal{V} = \mathrm{Var}(L \vee D) \cap \mathcal{V} = \mathrm{Var}(L \vee \overline{\Delta}(D)) \cap \mathcal{V}$, necessarily $\mathrm{Var}(C) \cap \mathcal{V} = \mathrm{Var}(C')$. Furthermore, since $C\nu_C = (L \vee D)\nu_D$ and $U_I \models \neg\Delta(D\nu_D)$, for every ground substitution $\sigma$ such that $dom(\sigma) \subseteq \mathcal{V}$,

$$U_I \cup \{C\nu_C\sigma\} \;\models\; U_I \cup \{(\overline{\Delta}(D) \vee L)\nu_D\sigma\} \;\models\; (\overline{\Delta}(D) \vee L)\delta\sigma \;\models\; C'\sigma.$$

The abstracted clause $C$ and the standard clause $C\nu_C$ are logically equivalent, hence $U_I \cup \{C\sigma\} \models C'\sigma$. We now prove that $C'$ is either in $S_I$, or is redundant in $S_I$.

If $C'$ is an $\mathcal{A}$-clause then necessarily $E'_1 = E'_2 = \overline{\Delta}(D) = \Box$ and $C' = L\delta$. But $C$ must also be an $\mathcal{A}$-clause in this case and since $S$ is $\mathcal{A}$-saturated, $C\nu_C$ is either in $S$ or is entailed by a subset of the $\mathcal{A}$-clauses in $S$; in both cases, $U_I \models C$. But $U_I \models \neg\Delta(D\nu_D)$, thus $U_I \models L\nu_D$ and $U_I \models L\delta$. Since $U_I$ contains all equalities between abducible constants that have the same interpretation under $I$, we conclude that $C'$ must be in $U_I$.

Now assume that $C'$ is not an $\mathcal{A}$-clause, and suppose that $C \notin S$. Let $\sigma$ be a ground substitution such that $dom(\sigma) \subseteq \mathcal{V}$. By hypothesis, $C$ is $\mathcal{A}$-redundant in $S$, hence there exists a set of clauses $T$ that consists of instances of abstracted clauses in $S_\nu$, such that $T \models C\sigma$ and for all $E \in T$, $E \prec_{\mathcal{A}} C\sigma\nu_C$. Let $T_I = \{\overline{\Delta}(E)_{\|I} \mid E \in T \wedge U_I \models \neg\Delta(E)\}$, then $T_I \subseteq S_I$, and

$$T_I \cup U_I \;\models\; T \cup U_I \;\models\; \{C\sigma\} \cup U_I \;\models\; C'\sigma,$$

because $\mathrm{Var}(C) \cap \mathcal{V} = \mathrm{Var}(C')$. Since $C'$ is not an $\mathcal{A}$-clause, by definition of $\prec$, for all $E \in T_I \cup U_I$, $E \prec C'$. Consider the case where $C \in S$. If $U_I \models C$, then $U_I \models C'$ and again $C'$ must be redundant. Otherwise, $U_I \models \neg\Delta(C\nu_C)$, and $S_I$ must contain $\overline{\Delta}(C\nu_C)_{\|I}$. Since $\overline{\Delta}(C\nu_C)_{\|I} = \overline{\Delta}(C)\delta = \overline{\Delta}(L \vee D)\delta$, Equation 5 shows that $\overline{\Delta}(C\nu_C)_{\|I}$ subsumes $(\overline{\Delta}(D) \vee L)\delta = C'$, thus proving that $C'$ is subsumed by a clause in $S_I$. Therefore, $S_I$ is indeed saturated. ■

Since $S_I$ is saturated for the standard superposition calculus $\mathcal{SP}_{\prec}$ and contains no occurrence of the empty clause, we deduce that it is satisfiable. We obtain the main result of this section:

**Theorem 31** *Let $S$ be a set of abstracted clauses that is $\mathcal{V}_\mathcal{A}$-stable and contains no variable-eligible clause. If $S$ is $\mathcal{A}$-saturated and does not contain the empty clause, then $S$ is satisfiable.*

This theorem proves the refutational completeness of $\mathcal{SP}_\mathcal{A}$ together with contraction rules that eliminate $\mathcal{A}$-redundant clauses, for those sets of abstracted clauses $S$ whose saturation is guaranteed to meet the requirements of the theorem. The first two requirements are not restrictive: the abstraction of a set of standard clauses described right before Section 3.3 produces a set of abstracted and $\mathcal{V}_\mathcal{A}$-stable clauses, and the saturation of this set is guaranteed to only contain abstracted and $\mathcal{V}_\mathcal{A}$-stable clauses by Theorem 21. The fact that $S$ contains no variable-eligible clause cannot be imposed that easily, but such a condition is guaranteed if $S$ is variable-inactive, which is the case for many classes of clause sets of interest [2, 1].

Note that this completeness result is not – by itself – sufficient for our purpose, since our goal is not merely to test the satisfiability of clause sets but rather to generate flat consequences they logically entail. The next section shows how the calculus $\mathcal{SP}_\mathcal{A}$ can be employed to reach this goal.

## 5 A generation of explanations

We return to the problem of explaining why a set of clauses is satisfiable, and show how $\mathcal{SP}_\mathcal{A}$ can be used to generate explanations relating abducibles to one another. Given a satisfiable set of clauses $S'$, we denote by $\mathrm{I}_\mathcal{A}(S')$ the set of all $\mathcal{A}$-implicates of $S'$:

$$\mathrm{I}_\mathcal{A}(S') \stackrel{\text{def}}{=} \left\{C \text{ an } \mathcal{A}\text{-clause} \mid C \text{ is ground and } S' \models C\right\}.$$

It is clear that the all the information about abducibles constants that is entailed by $S'$ is contained in $\mathrm{I}_\mathcal{A}('S)$. However this set can be very large and it contains a lot of non-pertinent information, for example all logical tautologies, or all instances of the equality axioms. It therefore does not seem reasonable to return this entire set to a user. Another solution could be to return a subset $T \subseteq \mathrm{I}_\mathcal{A}(S')$ such that $T \vdash \mathrm{I}_\mathcal{A}(S')$, but again, such a set might be large and contain unnecessary information.

**Example 32** Consider the set $S' = \{f(a) \not\simeq f(c),\ g(b) \simeq c,\ g(y) \not\simeq c\}$, where $\mathcal{A} = \{a, b, c\}$. This set is satisfiable and $\mathrm{I}_\mathcal{A}(S)$ contains the $\mathcal{A}$-clauses $c \not\simeq a$ and $a \not\simeq b \vee c \not\simeq b$, and since one cannot derive into the other using $\mathcal{SP}$, they should both be in $T$. But the latter is a logical consequence of the former and may not be as useful to output.

The solution we choose is to return a (subsumption-minimal) subset $T' \subseteq \mathrm{I}_\mathcal{A}(S')$ satisfying the following property: for all $C \in \mathrm{I}_\mathcal{A}(S')$ that is not a tautology, there exists a clause $C' \in T'$ such that $C' \models C$. The clauses in $T'$ are the *prime implicates* of $S'$. The notion of prime implicates plays a central rôle in many applications of computer science and artificial intelligence, and several approaches have been proposed for computing the prime implicates of a given propositional formula (see, e.g., [9]). Some extensions to first-order logic have also been considered, such as, e.g., [11]. In what follows, we define an algorithm that computes prime implicates for sets of flat equational clauses.

It turns out that $\mathcal{SP}_\mathcal{A}$ cannot be used to determine the set $T'$. For instance, if $S' = \{a \simeq b, c \not\simeq d\}$, then the clause $a \not\simeq c \vee b \not\simeq d$ must be in $\mathrm{I}_\mathcal{A}(S')$. Since it is subsumed by no clause in $\mathrm{I}_\mathcal{A}(S')$ but itself, it must also be in $T'$, but no $\mathcal{SP}_\mathcal{A}$-inference rule (or $\mathcal{SP}$-inference rule for that matter) can be applied to $S'$ to generate such a clause. In the sequel, we will show how, starting with a set of $\mathcal{A}$-clauses that logically entails $\mathrm{I}_\mathcal{A}(S')$, it is possible to generate a set $T'$ using the *Resolution calculus*, denoted by $\mathcal{R}$ (we refer the reader to [10] for details on the Resolution calculus). From now on, $S'$ denotes a satisfiable set of standard clauses, and $S$ is a set of abstracted clauses such that $S_\nu = S'$. Thus, $S$ and $S'$ are equivalent. The first step towards this construction is the definition of a set of $\mathcal{A}$-clauses that logically entails $\mathrm{I}_\mathcal{A}(S')$. The (finite) set of all $\mathcal{A}$-clauses in the saturated set generated from $S$ using $\mathcal{SP}_\mathcal{A}$ will satisfy this requirement.

**Definition 33** We denote by $T_\infty$ the set of $\mathcal{A}$-clauses in the $\mathcal{A}$-saturated set generated from $S$ by $\mathcal{SP}_\mathcal{A}$. $\Diamond$

The key result that makes the generation of $\mathcal{A}$-implicates possible is that all the $\mathcal{A}$-clauses that are entailed by $S$ are actually logical consequences of $T_\infty$:

**Proposition 34** $T_\infty \models \mathrm{I}_\mathcal{A}(S')$.

Proof. Let $C \in \mathrm{I}_\mathcal{A}(S')$. Since $S' \cup \neg C$ is unsatisfiable by hypothesis, so is $S \cup \neg C$, and there exists an $\mathcal{SP}_\mathcal{A}$-refutation of this set. Since any $\mathcal{SP}_\mathcal{A}$-inference involving an $\mathcal{A}$-clause as a premise must actually have all its premises that are $\mathcal{A}$-clauses by Theorem 21, it is possible to extract from the $\mathcal{SP}_\mathcal{A}$-refutation of $S \cup \neg C$ an $\mathcal{SP}_\mathcal{A}$-refutation of $T_\infty \cup \neg C$, hence the result. ∎

Recall that this result does *not* hold for the standard superposition calculus: for instance $a \not\simeq b$ is a logical consequence of $f(a) \not\simeq f(b)$ but no ground, flat clause implying $a \not\simeq b$ can be derived from $f(a) \not\simeq f(b)$. This shows the interest of the calculus $\mathcal{SP}_\mathcal{A}$. Note that since $T_\infty \subseteq \mathrm{I}_\mathcal{A}(S_\nu)$, both sets are actually equivalent. Let $Eq$ be the set of axioms stating that $\simeq$ is an equivalence relation[4]: $Eq = \{x \simeq x,\ x \not\simeq y \vee y \simeq x,\ x \not\simeq y \vee y \not\simeq z \vee x \simeq z\}$, and let $Eq_\mathcal{A}$ be the set consisting of all instantiations of the axioms in $Eq$ by the elements in $\mathcal{A}$. The result we show is that the $\mathcal{R}$-closure of the set $T_\infty \cup Eq_\mathcal{A}$ satisfies the requirements for the set of $\mathcal{A}$-clauses that is searched for. The proof is based on the following property:

**Lemma 35** *Given a set $S$, a clause $L \vee C$ and a substitution $\sigma$ such that $L' = L\sigma$ is ground, let $\delta$ be an $\mathcal{R}$-derivation from $S \uplus \{C\sigma\}$ of a clause $D$. Then there exists an $\mathcal{R}$-derivation $\delta'$ from $S \cup \{L \vee C\}$ of a clause $D'$ such that there exists an $r \geq 0$ and a substitution $\mu$ verifying $D'\mu = L'^r \vee D$, where the notation $L'^r$ means literal $L'$ is repeated $r$ times.*

[4]There will be no need to consider the congruence axiom, since all the clauses in $T_\infty$ only contain constants.

Proof. The result is proved by induction on the length of $\delta$. If $\delta = (D)$, then necessarily $D \in S \cup \{C\sigma\}$. It is simple to verify that the result holds when $D \in S$ and when $D = C\sigma$.

Assume that there are clauses $C_i = M \vee E_i$ and $C_j = N \vee E_j$ in $\delta$ such that $D = (E_i \vee E_j)\theta$, where $\theta$ is the mgu of $M, N$. By the induction hypothesis, there are clauses $C'_i$ and $C'_j$ generated from $S \cup \{L \vee C\}$ such that:

- there exists $r_i \geq 0$ and a substitution $\mu_i$ such that $C'_i\mu_i = L'^{r_i} \vee M \vee E_i$, and
- there exists $r_j \geq 0$ and a substitution $\mu_j$ such that $C'_j\mu_j = L'^{r_j} \vee N \vee E_j$.

Thus we have $C'_i = M' \vee F'_i$ where $M'\mu_i = M$ and $C'_j = N' \vee F'_j$ where $N'\mu_j = N$. Let $\mu_0 = \mu_i \cup \mu_j$. Since $M'\mu_0\theta = N'\mu_0\theta$, necessarily $M'$ and $N'$ are unifiable with mgu $\theta'$, and there exists a substitution $\mu$ such that $\mu_0\theta = \theta'\mu$. The Resolution rule applied to $C'_i, C'_j$ generates the clause $D' = (F'_i \vee F'_j)\theta'$, and we have

$$D'\mu \;=\; (F'_i \vee F'_j)\theta'\mu \;=\; (F'_i \vee F'_j)\mu_0\theta \;=\; (L'^{(r_i+r_j)} \vee E_i \vee E_j)\theta \;=\; L'^{(r_i+r_j)} \vee D,$$

where the last equality comes from the fact that by hypothesis, $L'$ is ground. The proof when $D$ is generated by the Factorization rule is similar. ∎

This permits to prove the main result of the section:

**Theorem 36** *Let $T = T_\infty \cup Eq_{\mathcal{A}}$, and let $C$ be a non-tautological ground clause in $\mathrm{I}_{\mathcal{A}}(S)$. Then there is a derivation from $T$ of a clause $C'$ such that $C' \models C$.*

Proof. By Proposition 34, $T_\infty \models C$, hence, by refutational completeness of the Resolution calculus, there exists a refutation of $T \cup \neg C$. We prove the result by induction on the length of the refutation. If the refutation is of length 1 then necessarily $\Box \in T$, and it is clear that $\Box \models C$. Assume this refutation is of length $n \geq 2$, then

$$T \cup \neg C \;\vdash_{\mathcal{R}}\; T \cup \neg C \cup \{D\sigma\} \;\vdash_{\mathcal{R}}^{n-1}\; \Box.$$

By the induction hypothesis, there exists a derivation from $T \cup \{D\sigma\}$ of a clause $C'$ such that $C' \models C$. If $D\sigma \in T$, then the result trivially holds. Now assume that $D\sigma \notin T$, we consider three cases depending on the sets the premises come from.

- If both premises are in $T$, then obviously there exists a derivation from $T$ that generates $C'$, and the result holds.
- If $D\sigma$ is generated from premises in $\neg C$, then since all the elements in this set are unit clauses, $C$ has to be of the form $L \vee L^c \vee C''$, which is impossible since $C$ is assumed not to be a tautology.
- Otherwise, $D\sigma$ is generated by a Resolution inference on a clause $L \vee D$ in $T$ and a unit clause $L'^c$ in $\neg C$ which is ground. Thus, $\sigma$ is the mgu of $L, L'$ and $L\sigma = L'$ must be ground. Since $T \cup \{D\sigma\}$ generates the clause $C'$, by Lemma 35, $T$ generates a clause $D'$ such that there exists an $r \geq 0$ and a substitution $\mu$ verifying $D'\mu = L'^r \vee C'$. Since $L'$ is a literal in $C$, we deduce that $D' \models C$, hence the result. ∎

$\textsc{Explain}(S', \mathcal{A}) =$
$S := \textsc{Abstract}(S)$
$S := \mathcal{SP}_{\mathcal{A}}$-saturation$(S)$
$T_\infty := \{C \in S \mid C$ is an $\mathcal{A}$-clause$\}$
**return** $\mathcal{R}$-saturation$(T_\infty \cup Eq_{\mathcal{A}})$

Figure 3: Generation of a set of explanations

To summarize, given a set of clauses $S'$ that is satisfiable and a set of abducible constants $\mathcal{A}$, the simple algorithm in pseudo-code described in Figure 3 returns a set of clauses constructed over $\mathcal{A}$ that can be viewed as explanations why $S'$ is satisfiable. Note that $\mathcal{R}$-saturation can be performed on the fly: it is clear that it is not necessary to wait until $\mathcal{SP}_A$-saturation$(S)$ is computed to start generating the clauses in $\mathcal{R}$-saturation$(T_\infty \cup Eq_{\mathcal{A}})$. Thus even in case of non-termination, all the prime implicates can eventually be generated. After the set $\mathcal{R}$-saturation$(T_\infty \cup Eq_{\mathcal{A}})$ is computed, it is possible to remove from this set all the clauses that can be inferred from other prime implicates. This solution yields a more compact representation. However, this is possible only in case of termination, since the deleted clauses may be involved in the generation of other prime implicates. A termination result for $\mathcal{SP}_{\mathcal{A}}$ will be presented in the following section. By putting all the previous results together, we obtain the following theorem, stating the soundess and completeness of the procedure Explain.

**Theorem 37** *Let $S$ be a set of clauses. Every clause $C \in \textsc{Explain}(S', \mathcal{A})$ is an $\mathcal{A}$-implicate of S, and for every $\mathcal{A}$-implicate $C$ of $S$ that is not a tautology, there exists a clause $C' \in \textsc{Explain}(S', \mathcal{A})$ such that $C' \models C$.*

**Example 38** We return to the problem mentioned in the Introduction. After flattening, we get the following set of clauses:

| | | | |
|---|---|---|---|
| 1 | $\mathrm{select}(\mathrm{store}(x,z,v),z) \simeq v$ | 4 | $d_2 \simeq \mathrm{store}(d_1,j,c)$ |
| 2 | $z \simeq w \vee \mathrm{select}(\mathrm{store}(x,z,v),w) \simeq \mathrm{select}(x,w)$ | 5 | $d_3 \simeq \mathrm{store}(a,j,c)$ |
| 3 | $d_1 \simeq \mathrm{store}(a,i,b)$ | 6 | $d_4 \simeq \mathrm{store}(d_3,i,b)$ |
| 7 | $\mathrm{select}(d_2,k) \not\simeq \mathrm{select}(d_4,k)$ | | |

Assume that $\mathcal{A} = \{i,j,b,c\}$. Then Clauses $3, 4, 5, 6$ are abstracted as follows:

| | |
|---|---|
| $3'$ | $x' \not\simeq i \vee y' \not\simeq b \vee d_1 \simeq \mathrm{store}(a,x',y')$ |
| $4'$ | $x'' \not\simeq j \vee y'' \not\simeq c \vee d_2 \simeq \mathrm{store}(d_1,x'',y'')$ |
| $5'$ | $x'' \not\simeq j \vee y'' \not\simeq c \vee d_3 \simeq \mathrm{store}(a,x'',y'')$ |
| $6'$ | $x' \not\simeq i \vee y' \not\simeq b \vee d_4 \simeq \mathrm{store}(d_3,x',y')$ |

$\mathcal{SP}_{\mathcal{A}}$ generates the following clauses[5]:

| | | |
|---|---|---|
| 8 | $x' \not\simeq i \vee w \simeq x' \vee \mathrm{select}(d_1, w) \simeq \mathrm{select}(a, w)$ | (3′,2) |
| 9 | $x'' \not\simeq j \vee w \simeq x'' \vee \mathrm{select}(d_2, w) \simeq \mathrm{select}(d_1, w)$ | (4′,2) |
| 10 | $x'' \not\simeq j \vee w \simeq x'' \vee \mathrm{select}(d_3, w) \simeq \mathrm{select}(a, w)$ | (5′,2) |
| 11 | $x' \not\simeq i \vee w \simeq x' \vee \mathrm{select}(d_4, w) \simeq \mathrm{select}(d_3, w)$ | (6′,2) |
| 12 | $x' \not\simeq i \vee y' \not\simeq b \vee \mathrm{select}(d_1, x') \simeq y'$ | (3′,1) |
| 13 | $x'' \not\simeq j \vee y'' \not\simeq c \vee \mathrm{select}(d_2, x'') \simeq y''$ | (4′,1) |
| 14 | $x'' \not\simeq j \vee y'' \not\simeq c \vee \mathrm{select}(d_3, x'') \simeq y''$ | (5′,1) |
| 16 | $x' \not\simeq i \vee y' \not\simeq b \vee \mathrm{select}(d_4, x') \simeq y'$ | (6′,1) |
| 17 | $x' \not\simeq i \vee k \simeq x' \vee \mathrm{select}(d_2, k) \not\simeq \mathrm{select}(d_3, k)$ | (11, 7) |
| 18 | $x' \not\simeq i \vee k \simeq x' \vee x'' \not\simeq j \vee k \simeq x'' \vee \mathrm{select}(d_2, k) \not\simeq \mathrm{select}(a, k)$ | (10, 17) |
| 19 | $x' \not\simeq i \vee k \simeq x' \vee x'' \not\simeq j \vee k \simeq x'' \vee \mathrm{select}(d_1, k) \not\simeq \mathrm{select}(a, k)$ | (9, 18) |
| 20 | $x' \not\simeq i \vee x'' \not\simeq j \vee k \simeq x' \vee k \simeq x''$ | (8,19) |
| 21 | $x' \not\simeq i \vee x'' \not\simeq j \vee k \simeq x' \vee \mathrm{select}(d_2, k) \not\simeq \mathrm{select}(d_4, x'')$ | (20,7) |
| 22 | $x' \not\simeq i \vee x'' \not\simeq j \vee k \simeq x' \vee x'' \simeq x' \vee \mathrm{select}(d_2, k) \not\simeq \mathrm{select}(d_3, x'')$ | (11,21) |
| 23 | $x' \not\simeq i \vee x'' \not\simeq j \vee y'' \not\simeq c \vee k \simeq x' \vee x'' \simeq x' \vee \mathrm{select}(d_2, k) \not\simeq y''$ | (14,22) |
| 24 | $x' \not\simeq i \vee x'' \not\simeq j \vee y'' \not\simeq c \vee k \simeq x' \vee x'' \simeq x' \vee \mathrm{select}(d_2, x'') \not\simeq y''$ | (20,23) |
| 25 | $x' \not\simeq i \vee x'' \not\simeq j \vee k \simeq x' \vee x'' \simeq x'$ | (13,24) |
| 26 | $x' \not\simeq i \vee x'' \not\simeq j \vee x'' \simeq x' \vee \mathrm{select}(d_2, k) \not\simeq \mathrm{select}(d_4, x')$ | (25,7) |
| 27 | $x' \not\simeq i \vee x'' \not\simeq j \vee y' \not\simeq b \vee x'' \simeq x' \vee \mathrm{select}(d_2, k) \not\simeq y'$ | (16,26) |
| 28 | $x' \not\simeq i \vee x'' \not\simeq j \vee y' \not\simeq b \vee x'' \simeq x' \vee \mathrm{select}(d_2, x') \not\simeq y'$ | (25,27) |
| 29 | $x' \not\simeq i \vee x'' \not\simeq j \vee y' \not\simeq b \vee x'' \simeq x' \vee \mathrm{select}(d_1, x') \not\simeq y'$ | (9,28) |
| 30 | $i \simeq j$ | (12,29) |
| 31 | $x' \not\simeq i \vee x'' \not\simeq j \vee x' \not\simeq x'' \vee k \simeq x'$ | (20) |
| 33 | $x' \not\simeq i \vee x'' \not\simeq j \vee x' \not\simeq x'' \vee \mathrm{select}(d_2, k) \not\simeq \mathrm{select}(d_4, x')$ | (31,7) |
| 34 | $x' \not\simeq i \vee x'' \not\simeq j \vee x' \not\simeq x'' \vee y' \not\simeq b \vee \mathrm{select}(d_2, k) \not\simeq y'$ | (16,34) |
| 35 | $x' \not\simeq i \vee x'' \not\simeq j \vee x' \not\simeq x'' \vee y' \not\simeq b \vee \mathrm{select}(d_2, x') \not\simeq y'$ | (31,34) |
| 36 | $i \not\simeq j \vee b \not\simeq c$ | (13,35) |

By Resolution, from 30 and 36, we get $c \not\simeq b$, which subsumes 36. We obtain the $A$-implicates $\{i \simeq i, b \not\simeq c\}$, yielding the explanation $i \not\simeq j \vee b \simeq c$.

## 6 A termination result for $\mathcal{SP}_{\mathcal{A}}$

In this section we will prove a result that relates the termination of $\mathcal{SP}$ on a set of standard clauses $S$ to the termination of $\mathcal{SP}_{\mathcal{A}}$ on an abstracted version of $S$. This shows that many existing results about the termination of the superposition calculus for subclasses of first-order logic carry over to $\mathcal{SP}_{\mathcal{A}}$.

We introduce a way to relate standard and abstracted terms by defining a so-called relation of $\mathcal{A}$-relaxation. This relation will be used afterwards to relate the forms of the clauses generated by $\mathcal{SP}$-inferences and those generated by $\mathcal{SP}_{\mathcal{A}}$-inferences in a more precise manner.

**Definition 39** The relation of *$\mathcal{A}$-relaxation* relates an abstracted term $t$ to a standard one $t'$ and is defined as follows: $t \trianglelefteq_{\mathcal{A}} t'$ if and only if $t\gamma_0 = t'_{\downarrow\mathcal{A}}$.

[5]For readability we simply drop irrelevant disequations, i.e. $x \not\simeq a \vee C$ is replaced by $C$ if $x$ does not occur in $C$ and $x \not\simeq a \vee x' \not\simeq a \vee C$ is replaced by $x \not\simeq a \vee C\{x' \mapsto x\}$.

Given an abstracted clause $C$ and a standard clause $C'$, we write $C \trianglelefteq_{\mathcal{A}} C'$ if and only if $\overline{\Delta}(C\gamma_0) = \overline{\Delta}(C'_{\downarrow\mathcal{A}})$. This relation is extended to sets of clauses in a straightforward manner. $\diamondsuit$

**Example 40** Assume $\mathcal{A} = \{a, b\}$, let $\mathrm{C} = x \not\simeq a \vee a \simeq b \vee f(x, x, d) \simeq g(y) \vee g(y) \simeq d$ and $C' = f(a, b, d) \simeq g(b) \vee g(a) \simeq d$. Then $C \trianglelefteq_{\mathcal{A}} C'$.

Note that all $\mathcal{A}$-literals are discarded when comparing clauses with the relation $\trianglelefteq_{\mathcal{A}}$.

**Lemma 41** *Let $C$ be an abstracted clause such that $\overline{\Delta}(C) \neq \Box$, let $C'$ be an $\mathcal{A}$-reduced clause and assume $C \trianglelefteq_{\mathcal{A}} C'$. Let $L = t \bowtie s$ be an $\mathcal{A}$-maximal literal in $C$ and $t$ be an $\mathcal{A}$-maximal term in $L$. If $L' = t' \bowtie s'$ is a literal in $C'$ such that $L \trianglelefteq_{\mathcal{A}} L'$, then $L'$ is a $\prec$-maximal literal in $C'$ and $t'$ is a $\prec$-maximal term in $L'$.*

Proof. Since $\overline{\Delta}(C) \neq \Box$, by definition of $\prec$, $L$ must be a literal in $\overline{\Delta}(C)$, so that $L'$ is well-defined. Furthermore, since $C'$ is $\mathcal{A}$-reduced, $L\gamma_0 = L'_{\downarrow\mathcal{A}} = L'$ by definition. Assume that $L'$ is not maximal in $C'$, and consider a literal $M'$ in $C'$ such that $L' \prec M'$. Necessarily, $M'$ occurs in $\overline{\Delta}(C')$, and there must exist a literal $M$ in $\overline{\Delta}(C)$ such that $M \trianglelefteq_{\mathcal{A}} M'$. Again by definition, $M\gamma_0 = M'_{\downarrow\mathcal{A}} = M'$, thus, by definition of $\prec_{\mathcal{A}}$, $L \prec_{\mathcal{A}} M$. This proves that $L$ cannot be $\mathcal{A}$-maximal, a contradiction. The proof that $t'$ is a maximal term in $L'$ is similar. ■

Lemma 44 can be viewed as a dual version of Lemma 22, where abstracted and standard terms are switched. We state some preliminary properties before proving the lemma.

**Proposition 42** *If $\sigma$ and $\sigma'$ are $\mathcal{A}$-compliant substitutions then $(\sigma\gamma_0)(\sigma'\gamma_0) = \sigma\sigma'\gamma_0$.*

**Proposition 43** *If $t \trianglelefteq_{\mathcal{A}} t'$ and $\sigma$ is $\mathcal{A}$-compliant, then $t'\sigma\gamma_0 = t'(\sigma\gamma_0)_{|\mathcal{V}}$ and $t\sigma \trianglelefteq_{\mathcal{A}} t'\sigma\gamma_0$.*

Proof. It is clear that $t'\sigma\gamma_0 = t'(\sigma\gamma_0)_{|\mathcal{V}}$, since all variables in $t'$ must be in $\mathcal{V}$. By definition $t\gamma_0 = t'_{\downarrow\mathcal{A}}$ and by Proposition 42 $t\gamma_0\sigma\gamma_0 = t\sigma\gamma_0$. Since $(t'_{\downarrow\mathcal{A}})\sigma\gamma_0 = (t'\sigma\gamma_0)_{\downarrow\mathcal{A}}$, we have the result. ■

**Lemma 44** *Let $t, s$ be abstracted terms, $t' = t\gamma_0$ and $s' = s\gamma_0$. If $t, s$ are $\mathcal{A}$-unifiable with mgu $\mu$, then $t', s'$ are unifiable with mgu $(\mu\gamma_0)_{|\mathcal{V}}$.*

Proof. The result is proved by induction on the size of $t\mu$. If one of the terms is in $\mathcal{V}_{\mathcal{A}}$ or in $\mathcal{V}$, then it is simple to verify that the result holds. Assume that $t = f(t_1, \ldots, t_n)$, and $s = f(s_1, \ldots, s_n)$, and let $t' = f(t'_1, \ldots, t'_n)$ and $s' = f(s'_1, \ldots, s'_n)$.

We let $\pi_0 \stackrel{\text{def}}{=} \mathrm{id}$, and for $i = 1, \ldots, n$, $\mu_i$ denotes the mgu of $t_i\pi_{i-1}, s_i\pi_{i-1}$ and $\pi_i \stackrel{\text{def}}{=} \pi_{i-1}\mu_i$. Since $t$ and $s$ are $\mathcal{A}$-unifiable, for all $i = 1, \ldots, n$, $t_i\pi_{i-1}$ and $s_i\pi_{i-1}$ are $\mathcal{A}$-unifiable and $\mu_i$ is $\mathcal{A}$-compliant. Thus, so is $\pi_i$ by Proposition 9, and $\mu = \pi_n$.

Let $\pi'_0 = \mathrm{id}$, for all $i = 1, \ldots, n$, let $\mu'_i$ denote the mgu of $t'_i\pi'_{i-1}$ and $s'_i\pi'_{i-1}$, and let $\pi'_i = \pi'_{i-1}\mu'_i$. We show by induction on $i$ that $t'_i\pi'_{i-1}$ and $s'_i\pi'_{i-1}$ are unifiable and that $\pi'_i = (\pi_i\gamma_0)_{|\mathcal{V}}$. This will prove that $t'$ and $s'$ are unifiable with mgu $\mu' \stackrel{\text{def}}{=} \pi'_n = (\pi_n\gamma_0)_{|\mathcal{V}} = (\mu\gamma_0)_{|\mathcal{V}}$.

Assume this result holds for $i-1$, and consider the terms $t'_i\pi'_{i-1}$ and $s'_i\pi'_{i-1}$. Since $t_i \trianglelefteq_{\mathcal{A}} t'_i$ and $s_i \trianglelefteq_{\mathcal{A}} s'_i$, we deduce that $t_i\pi_{i-1} \trianglelefteq_{\mathcal{A}} t'_i\pi'_{i-1}$ and $s_i\pi_{i-1} \trianglelefteq_{\mathcal{A}} s'_i\pi'_{i-1}$ by Proposition 43. Since the size of $t_i\pi_{i-1}$ is strictly less than that of $t\mu$, we may apply the induction hypothesis to conclude that $t'_i\pi'_{i-1}$ and $s'_i\pi'_{i-1}$ are unifiable with mgu $\mu'_i = (\mu_i\gamma_0)_{|\mathcal{V}}$. Therefore $\pi'_i$ is well-defined. The fact that no variable in $\mathcal{V}_{\mathcal{A}}$ occurs in the codomain of $\pi'_{i-1} = \pi_{i-1}\gamma_0$ together with Proposition 42 permits to verify that,

$$\pi'_i \;=\; \pi'_{i-1}\mu'_i \;=\; (\pi_{i-1}\gamma_0)_{|\mathcal{V}}(\mu_i\gamma_0)_{|\mathcal{V}} \;=\; (\pi_{i-1}\gamma_0\mu_i\gamma_0)_{|\mathcal{V}} \;=\; (\pi_i\gamma_0)_{|\mathcal{V}},$$

hence the result. ∎

**Lemma 45** *Let $C_1, C_2$ be abstracted clauses that are not variable-eligible, and $C'_1, C'_2$ be $\mathcal{A}$-reduced clauses such that $C_1 \trianglelefteq_{\mathcal{A}} C'_1$ and $C_2 \trianglelefteq_{\mathcal{A}} C'_2$. Assume further that neither $C_1$ nor $C_2$ is an $\mathcal{A}$-clause. If $C_1, C_2 \vdash^{\mathcal{A}} C$, then $C'_1, C'_2 \vdash C'$ and $C \trianglelefteq_{\mathcal{A}} C'$.*

Proof. We show the result for a paramodulation or superposition inference, the other cases are similar. Let

$$\begin{array}{rcl} C_1 & = & u \simeq v \vee D_1, \\ C_2 & = & t[w] \bowtie s \vee D_2, \\ C & = & (t[v] \bowtie s \vee D_1 \vee D_2)\sigma, \end{array}$$

where $\sigma$ is the mgu of $u$ and $w$. Then by Lemma 41, we have

$$\begin{array}{rcl} C'_1 & = & u' \simeq v' \vee D'_1, \\ C'_2 & = & t'[w'] \bowtie s' \vee D'_2, \end{array}$$

where $u \trianglelefteq_{\mathcal{A}} u'$ and $t[w] \trianglelefteq_{\mathcal{A}} t'[w']$. By Lemma 41, $u'$ (resp. $t'[w']$) is maximal. Thus[6] $C'_1, C'_2 \vdash C'$, where $C' = (t'[v'] \bowtie s' \vee D'_1 \vee D'_2)\sigma'$ and $\sigma'$ is the mgu of $u'$ and $w'$. By Lemma 44, $\sigma' = (\sigma\gamma_0)_{|\mathcal{V}}$, and by Proposition 43, $C_1\sigma \trianglelefteq_{\mathcal{A}} C'_1\sigma'$ and $C_2\sigma \trianglelefteq_{\mathcal{A}} C'_2\sigma'$. It is then straightforward to verify that $C \trianglelefteq_{\mathcal{A}} C'$. ∎

We define a notion of redundancy that is meant to hold no matter what abducible constants occur in the clause under consideration.

**Definition 46** An $\mathcal{A}$-reduced clause $C'$ is P-*redundant* in an $\mathcal{A}$-reduced set of clauses $S'$ if for all sets of abstracted clauses $S$ such that $(S_\nu)_{\downarrow\mathcal{A}} \equiv S'$ and for every abstracted clause $D$ such that $(D\nu_D)_{\downarrow\mathcal{A}} \equiv C'$, clause $D$ is $\mathcal{A}$-redundant in $S$. An $\mathcal{A}$-reduced set of clauses $S'$ is P-*saturated* if every clause generated with premises in $S'$ either occurs in $S'$ or is P-redundant in $S'$. ◊

[6] The strict maximality conditions for binary inference rules were relaxed in the version of the superposition calculus presented in Figure 2 to allow the following inference.

This notion permits to eliminate clauses that are redundant in the usual sense and do not contain any abducible constant.

**Example 47** Assume $\mathcal{A} = \{a, b\}$ and $S' = \{f(c) \not\simeq f(d)\}$, then $C' = g(a, c) \simeq h(a) \vee f(c) \not\simeq f(d)$ is P-redundant in $S'$.

**Theorem 48** *Let $S'$ be a set of $\mathcal{A}$-reduced clauses, and let $T$ be the* P*-saturated set of clauses generated from $S'$. If $T$ is finite and $S$ is a set of abstracted clauses that is $\mathcal{V}_{\mathcal{A}}$-stable, variable-inactive and such that $S \trianglelefteq_{\mathcal{A}} S'$, then the set of non-redundant clauses generated from $S$ is finite.*

Proof. By Theorem 21, for all $n \geq 0$, every non-redundant clause $C$ generated from $S$ is $\mathcal{V}_{\mathcal{A}}$-stable and it cannot be variable-eligible; by Lemma 45, there is a clause $C'$ generated by a derivation from $S'$ such that $C \trianglelefteq_{\mathcal{A}} C'$. The clause $C'$ cannot be P-redundant because otherwise, by definition, $C$ would be $\mathcal{A}$-redundant. Thus $C' \in T$, and since the set $\{D \mid D \trianglelefteq_{\mathcal{A}} C\}$ is finite up to equivalence, we deduce that the set of non-redundant clauses generated from $S$ is also finite. ∎

Theorem 48 guarantees that $\mathcal{SP}_{\mathcal{A}}$ (and thus Explain) terminates on several classes of clause sets, in particular for clause sets related to SMT problems. The authors of [2] and [1] prove that sets of the form $\mathcal{T} \cup S$, where $\mathcal{T}$ is a theory and $S$ a set of ground unit clauses, generate finite saturated sets. This result is extended to clause sets of the form $\mathcal{T} \cup S'$, where $S'$ is an arbitrary set of ground clauses, in [6]. An inspection of the finiteness results of [2, 1, 6] shows that they hold not only for saturated sets but also for P-saturated sets, since the redundant clauses that are deleted are actually P-redundant: they do not contain any constants at all. Thus, $\mathcal{SP}_{\mathcal{A}}$ terminates for clause sets of the form $\mathcal{T} \cup S'$, where $S'$ is the abstraction of a set of ground clauses, and $\mathcal{T}$ is the axiomatization of the any of the following theories: records, integer offsets, possibly empty lists, arrays...

## 7 Discussion

We have presented a calculus that permits to reason on the relations involving abducible constants, that are logical consequences of a satisfiable set of clauses. These relations can be viewed as explanations of why the set is satisfiable, since any of their negations, when added to the original clause set, renders the latter unsatisfiable. We proved a completeness result for the calculus, along with a sufficient condition guaranteeing its termination on classes of clause sets, among which SMT problems in several theories of interest. To the best of our knowledge, this approach is novel and there are many interesting directions to explore. One first direction is to investigate what set of clauses can be considered as a *good* set of explanations, and determine what a good trade-off might be between a small set of explanations that may hide too many details, and a large set of explanations that may carry too much unnecessary information. Another line of work that is currently under investigation is the search for a more efficient way to generate explanations. Indeed, the saturation with the Resolution calculus in the presence of the equality axioms is not entirely satisfactory as far as efficiency is concerned, and it would be interesting to see how the calculus $\mathcal{SP}_{\mathcal{A}}$ can be enhanced to directly produce the required set of explanations. As far as other extensions are concerned, we plan to investigate how to extend these results to *abducible terms* and not only abducible constants, by allowing the occurrence of function symbols in $\mathcal{A}$. This would allow the derivation of non-ground explanations. Another possibility is to consider mixed literals, containing both abducible and non-abducible symbols. It would then be possible to generate explanations of the form $a \simeq 0$ without having to declare 0 as an abducible constant. We also plan on devising a calculus capable of efficiently generating explanations with abducibles interpreted in a particular theory, such as, e.g., arithmetic.